\documentclass[article,twocolumn]{revtex4}
\usepackage{graphicx}
\usepackage{color}
\newcommand{\cre}[1]{ #1^\dagger}
\newcommand{\vac}{|{\rm vac}\rangle}
\newcommand{\ket}[1]{\left| #1 \right\rangle}

\renewcommand{\Im}{\mathrm{Im}}
\renewcommand{\Re}{\mathrm{Re}}
\newcommand{\expect}[1]{\left\langle#1\right\rangle}

\begin{document}
\title{Phase shifts vs time delays: Sagnac and Hong-Ou-Mandel}
\author{S.J. van Enk}
\affiliation{Oregon Center for Optics, Department of Physics\\
University of Oregon,
Eugene, OR 97403}

\begin{abstract}
We point out that the Sagnac effect can be measured by means of the Hong-Ou-Mandel effect. The latter is not sensitive to phase shifts, 
and thus the Hong-Ou-Mandel Sagnac effect hinges on the fact that the Sagnac effect is, fundamentally, a time delay, not a phase shift.  
\end{abstract}
\maketitle
\section{Introduction}
The Sagnac effect refers to a difference in roundtrip times between co- and counter-propagating waves in a rotating ring
interferometer \cite{post,stedman,malykin,anandan,rizzi}. The effect is relativistic in the sense that a Galilean-invariant theory predicts a null effect \cite{dieks}. 
The effect is universal in at least two ways. It exists for any type of waves, including classical light waves and quantum de Broglie waves. Moreover, the time difference is independent of the propagation speed of the signal and hence is the same for any frequency (Fourier) component of the wave. In particular, for light waves propagating in a co-moving medium, the time difference does not depend on the medium's refractive index.

When nearly monochromatic CW laser waves are used in a Sagnac interferometer, the Sagnac time difference manifests itself as a phase shift, or, equivalently, as a fringe shift in the interference pattern (and in a ring cavity, as in commercial laser gyros, the effect is transformed into a frequency shift). 
In fact, in any interference experiment with monochromatic light,  time delays, phase shifts, and fringe shifts are all equivalent.
In general, though, they are not.
Here we point out that there are states of the (quantized) electromagnetic field whose interference behavior is very differently affected by time delays than by phase shifts. More precisely, there are states that are insensitive to constant (i.e., the same for each Fourier component) phase shifts, but that do sense constant time delays.
Such states then, can be used to directly test 
one fundamental aspect of the Sagnac effect, that it is a constant time shift, not a constant phase shift.
In particular, 
one example of an effect that is sensitive to time delays but not to phase shifts, is the Hong-Ou-Mandel effect \cite{HOM}. Interestingly, the inverse Hong-Ou-Mandel effect is sensitive to both, and is in fact twice as sensitive to phase shifts as classical light waves are (as was pointed out in Ref.~\cite{agarwal}, although there the effect was treated for monochromatic light, and hence did not distinguish between time shifts and phase shifts).

\section{Theory}
\subsection{Sagnac}
Consider a ring interferometer, enclosing an area $A$, rotating at a frequency $\Omega$ (see Figure 1). 
\begin{figure}[h]
 \includegraphics[width=.45\textwidth]{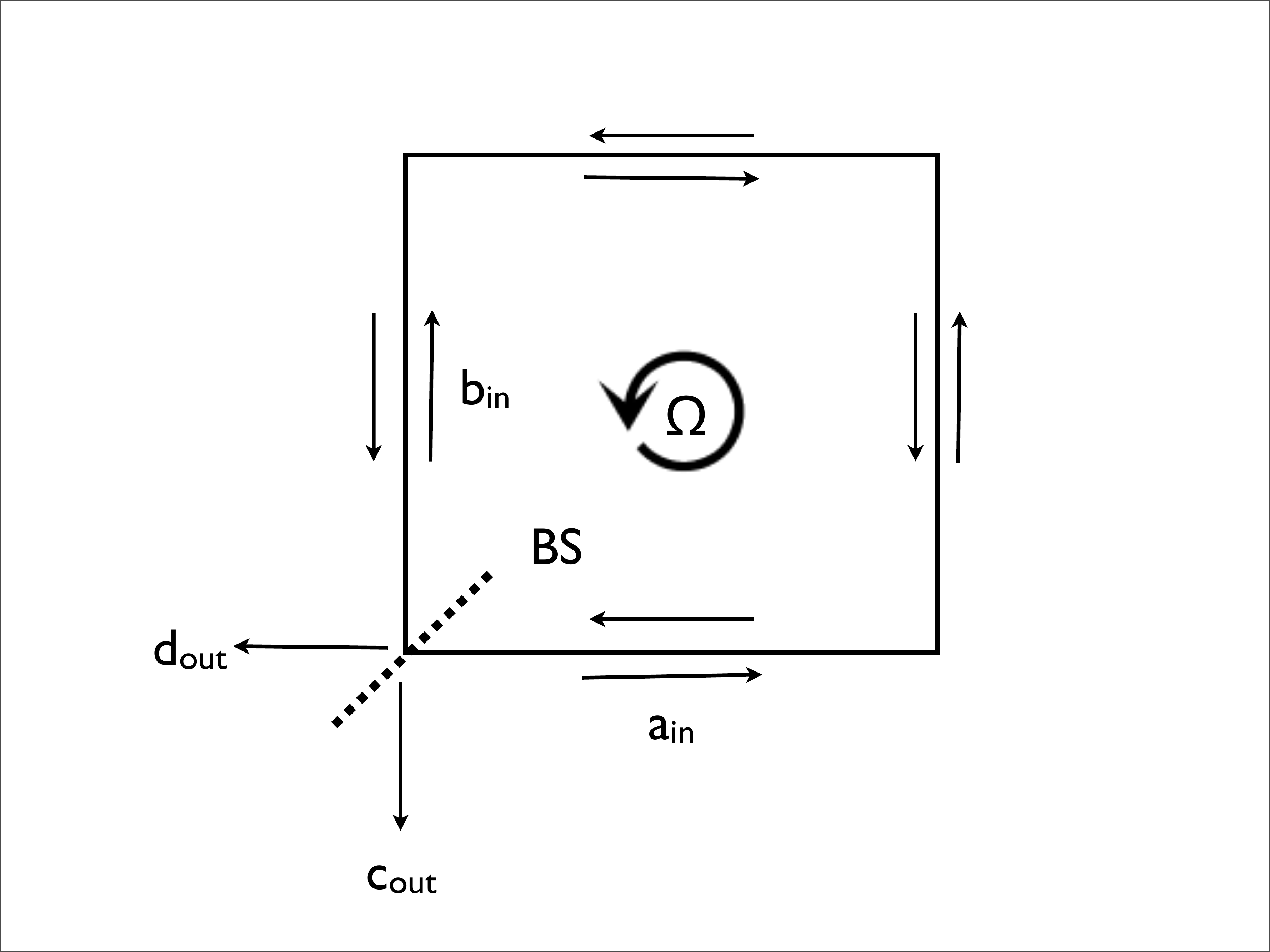}
  \caption{Sagnac interferometer rotating at a frequency $\Omega$. Light propagates through a co-rotating medium in both the counter-clock-wise  and clock-wise directions. The initial complex amplitudes of the two light waves at the source are indicated by $a_{{\rm in}}$ and
 $b_{{\rm in}}$, respectively. After one roundtrip the waves exit the interferometer  by impinging on a (co-rotating) 50/50 beamsplitter, resulting in output waves with amplitudes $c_{{\rm out}}$ and
 $d_{{\rm out}}$.}
\end{figure}
There are two sets of modes of interest, $a(\omega)$  and $b(\omega)$, co-propagating and counter-propagating, respectively. 
These modes propagate from their source to the detection point, whose locations as seen from the rotating frame of reference coincide.
The modes make a single roundtrip through a co-moving medium with refractive index $n(\omega)$ and take a time $\tau^{\pm}(\omega)$ to complete this roundtrip. 
We can distinguish two contributions to this total time: the first is present even when there is no rotation, and is determined by $n(\omega)$ and the distance traveled; this contribution is the same for co- and counter-propagating waves. The second term 
contains the Sagnac time delays, $\tau_S^{\pm}$, which depend on both $\omega$ and $\Omega$. Their difference, $\Delta \tau_S=\tau_S^+-\tau_S^-$ does {\em not} depend on the refractive index of the medium nor on $\omega$, and is given by
\begin{equation}
\Delta \tau_S=\frac{4A\Omega}{c^2}.
\end{equation}
More precisely, this is the time delay as measured from the lab frame, assumed to be an inertial frame of reference. There is an extra time dilation factor for the time delay as measured in the rotating frame of reference, but to first approximation (using $\Omega R/c$ as small expansion parameter, with $R$ the distance of the detector/source from the rotation axis) we can neglect this correction.

We assume the modes impinge on a 50/50 beamsplitter after one roundtrip, exit the ring interferometer, after which measurements are performed.
We can express the ``output'' amplitudes of the modes (after the beamsplitter, denoted by $c$ and $d$) in terms of the ``input'' amplitudes (at the source, denoted by $a$ and $b$ for co- and counter-propagating modes, resp.)  through 
\begin{eqnarray}\label{cdab}
c_{{\rm out}}(\omega)&=&\frac{a_{{\rm in}}(\omega)e^{i\omega\tau^+(\omega)}+ib_{{\rm in}}(\omega)e^{i\omega\tau^-(\omega)}}{\sqrt{2}},\nonumber\\
d_{{\rm out}}(\omega)&=&\frac{b_{{\rm in}}(\omega)e^{i\omega\tau^-(\omega)}+ia_{{\rm in}}(\omega)e^{i\omega\tau^+(\omega)}}{\sqrt{2}},
\end{eqnarray}
and measurements are performed on the output modes $c$ and $d$.
Assume now we measure the intensities of modes $c$ and $d$ (integrating over all frequencies $\omega$). We have\begin{eqnarray}\label{I}
\expect{I_{c,d}}&=&\frac12\int d\omega
\expect{I_a(\omega)}+\expect{I_b(\omega)}
\nonumber\\
&\pm&\Im\int d\omega
 e^{i\omega\Delta\tau_S}\expect{b^*_{{\rm in}}(\omega)a_{{\rm in}}(\omega)}.
\end{eqnarray}
The last term describes interference
caused by the Sagnac time difference.
If we replace the frequency-dependent phase factor $e^{i\omega\Delta\tau_S}$ by a frequency-independent phase factor $e^{i\phi}$ (i.e., a pure phase shift), the interference term is still nonzero in general. Thus, in general, classical waves are sensitive to both phase shifts and time delays (and the sensitivity is the same for monochromatic waves).

We can easily switch to a quantum description by promoting the complex amplitudes to annihilation operators (in the Heisenberg picture). In particular, if the input state of modes $a$ and $b$
is given by a general pure state
\begin{eqnarray}
\ket{\psi}_{{\rm in}}&=&\int d\omega
\int d\omega'\sum_{n,m} a_{nm}(\omega,\omega')\nonumber\\
&\times&(\cre{a}_{{\rm in}}(\omega))^n(\cre{b}_{{\rm in}}(\omega'))^m\vac,
\end{eqnarray}
then the output state of modes $c$ and $d$ can be found by inverting and taking the hermitian conjugate of the relations (\ref{cdab}), which results in
\begin{eqnarray}
\ket{\psi}_{{\rm out}}&=&\int d\omega\int d\omega'\sum_{n,m} a_{nm}(\omega,\omega')\nonumber\\
&\times&e^{in\omega\tau^+(\omega)}\left(\frac{\cre{c}_{{\rm out}}(\omega)+i\cre{d}_{{\rm out}}(\omega)}{\sqrt{2}}\right)^n\nonumber\\
&\times&e^{im\omega'\tau^-(\omega')}\left(\frac{\cre{d}_{{\rm out}}(\omega')+i
\cre{c}_{{\rm out}}(\omega')}{\sqrt{2}}\right)^m\vac,
\end{eqnarray}
where the invariance of the vacuum state $\vac$ under the (photon-number-preserving) unitary evolution was used.

It is easy to see what sort of input states are insensitive to {\em constant} (differential) phase
shifts, $a_{{\rm in}}(\omega)\mapsto
\exp(i\phi_a)a_{{\rm in}}(\omega)$ and
$b_{{\rm in}}(\omega)\mapsto
\exp(i\phi_b)b_{{\rm in}}(\omega)$, namely, 
states with definite photon numbers in both input modes $a_{{\rm in}}$ and $b_{{\rm in}}$, as there is only a physically irrelevant overall phase shift of the input (and output) state. But such states are, in general, sensitive to time delays (except the vacuum, of course).
\subsection{Hong-Ou-Mandel}
The prime example of a state that is sensitive to time delays---but not to constant phase shifts--- in exactly this same setup is the state with two spectrally identical photons, one in each input mode, as this is the state that features in the Hong-Ou-Mandel effect measuring that very time delay.
Indeed, any state of the form
\begin{equation}
\ket{\Psi}_{{\rm in}}=\int d\omega \int d\omega' \phi(\omega)\phi(\omega')
\cre{a}_{{\rm in}}(\omega)\cre{b}_{{\rm in}}(\omega')\vac,
\end{equation}
(with normalization $\int d\omega|\phi(\omega)|^2=1$)
leads to an output state with the property that at zero time delay one never finds 
one photon in each output mode, as the last line in the following equation shows:  
\begin{eqnarray}\label{HOM}
\ket{\Psi}_{{\rm out}}&=&
\frac i2\int d\omega\int d\omega'
\phi(\omega)\phi(\omega') e^{i\omega\tau^+(\omega)+i\omega'\tau^-(\omega')}\nonumber\\
&\times&
\left[\cre{c}_{{\rm out}}(\omega)
\cre{c}_{{\rm out}}(\omega')
+
\cre{d}_{{\rm out}}(\omega)
\cre{d}_{{\rm out}}(\omega')
\right.
\nonumber\\
&+&\left.
\left(1-e^{i(\omega'-\omega)\Delta \tau_S}\right)
\cre{c}_{{\rm out}}(\omega)\cre{d}_{{\rm out}}(\omega')\right]\vac.
\end{eqnarray}
Thus, the ($\omega$-independent) time delay $\Delta\tau_S$ can be measured directly by 
measuring coincidence counts between the two output modes (where, to mention the same point once more, an $\omega$-independent phase shift would give no such effect).

If we define
\begin{equation}\label{K}
K=\int d\omega |\phi(\omega)|^2
e^{-i\omega\Delta\tau_S},
\end{equation}
then the probability to detect one photon in each output mode is
\begin{equation}
P_{11}=\frac 12\left(1-|K|^2\right).
\end{equation}

\subsection{Inverse Hong-Ou-Mandel}
For completeness  let us analyze what happens when we use the output of a standard (i.e., in an inertial frame of reference) Hong-Ou-Mandel experiment as input to the Sagnac interferometer. That is, assume our input state has the form
\begin{eqnarray}
\ket{\Phi}_{{\rm in}}&=&
\int d\omega \int d\omega' \phi(\omega)\phi(\omega')\nonumber\\
&\times&
\left[\cre{a}_{{\rm in}}(\omega)
\cre{a}_{{\rm in}}(\omega')
+\cre{b}_{{\rm in}}(\omega)\cre{b}_{{\rm in}}(\omega')\right]\vac.
\end{eqnarray}
Then the output state will be 
\begin{eqnarray}\label{iHOM}
\ket{\Phi}_{{\rm out}}&=&
\int d\omega\int d\omega'
\phi(\omega)\phi(\omega')e^{i\omega\tau^+(\omega)+i\omega'\tau^+(\omega')} \nonumber\\
&\times&\Big[i\left\{1+e^{-i(\omega+\omega')\Delta\tau_S}\right\}\cre{c}_{{\rm out}}(\omega)
\cre{d}_{{\rm out}}(\omega')
\nonumber\\
&+&
\frac 12\left\{1-e^{-i(\omega+\omega')\Delta\tau_S}\right\}\nonumber\\&\times&\left(\cre{c}_{{\rm out}}(\omega)\cre{c}_{{\rm out}}(\omega')-
\cre{d}_{{\rm out}}(\omega)\cre{d}_{{\rm out}}(\omega')
\right)\Big]\vac.
\end{eqnarray}
The last term vanishes when $\Delta\tau_S=0$. The time delay can be measured now by counting the cases where both photons appear in one and the same output mode. The phase factor in the last line
now depends on the {\em sum} of the two (dummy) frequencies $\omega$ and $\omega'$, where the {\em difference} appeared in the last line of (\ref{HOM}): this is why the Hong-Ou-Mandel effect is not sensitive to constant phase shifts, but the inverse Hong-Ou-Mandel effect is twice as sensitive to phase shifts as classical light waves are [recall that in the classical case there appears just a phase factor $e^{i\omega\Delta \tau_S}$, see Eq.~(\ref{I})].

More precisely, in terms of the quantity $K$, defined in (\ref{K}), the probability to find
two photons in the same mode (either $c$ or $d$) is
\begin{equation}
P_{20}=\frac 12\left(1-\Re(K^2)\right).
\end{equation}
We always have $P_{20}\geq P_{11}$ with equality only if $K$ is real.
 
 \section{Discussion}
 When contemplating experimental implementation of the Hong-Ou-Mandel experiment on a rotating platform, there are a number of issues that spring to mind. Fortunately, some of those problems have been dealt with successfully in the only {\em single}-photon implementation of the Sagnac effect  so far \cite{berto}. The main features of that experiment was that a fiber setup was used, with the single-photon source (at 1550 nm) placed {\em outside} the rotating platform (any Doppler shifts can be neglected to first-order approximation).  
By winding the fiber loop many times around the platform, the Sagnac effect, being proportional to the area enclosed, is enhanced by the winding number, so that a small value of $\Omega$ is tolerable. In fact, a very clean fringe was observed with a visibility of more than 99\%, and the maximum time delay was about 100ps,  a time delay easily measurable with the Hong-Ou-Mandel effect.
(Remembering the present context, note this single-photon experiment was sensitive to both constant phase shifts and time delays. The input state used was of the form
\begin{eqnarray}
\ket{\psi'}_{{\rm in}}=\int d\omega
\phi'(\omega)
[\cre{a}_{{\rm in}}(\omega)+\cre{b}_{{\rm in}}(\omega)]\vac,
\end{eqnarray}
idealizing it as pure input state.)

The Hong-Ou-Mandel effect is more complicated, as it needs two (spectrally) identical photons (More precisely, in the rotating frame of reference they need to be identical spectrally. In a frame of reference relative to which the beamsplitter is moving, perfect interference takes place for spectrally different wave packets, thanks to the Doppler effect (see \cite{raymer})). In practice, of course, the two photons will be different. That will mean there will always be coincidence counts in the output. Similarly, a beamsplitter is never exactly 50/50, and that leads to coincidence counts as well, even with identical photons as input.
But these effects are all independent of the rotation rate $\Omega$. Performing a control experiment at zero rotation rate provides, therefore, a standard solution, by allowing one to subtract off the effects of imperfections.

Finally, creating a state of the form
$\ket{\Phi}_{{\rm in}}$ is in principle merely just as complicated as creating a state of the form
$\ket{\Psi}_{{\rm in}}$, as the former 
is created from the latter, by impinging it on a 50/50 beamsplitter.
In the present context, it has to be added that that 50/50 beamsplitter should be at rest in an inertial frame. This will make sure all inevitable imperfections in the input state will again be independent of the rotation rate $\Omega$.

\end{document}